\RequirePackage[displaymath]{lineno}
\documentclass[twocolumn,aps,prl,showpacs,superscriptaddress,floatfix,nofootinbib]{revtex4-2}
\usepackage{newtxtext,newtxmath,booktabs,siunitx}
\usepackage{dcolumn}
\usepackage{bm}
\usepackage{float}
\usepackage{ulem}
\usepackage[]{graphicx}  
\usepackage{booktabs}
\usepackage{tabularx}
\usepackage{amsmath}
\usepackage{color}
\usepackage{multirow}
\usepackage{comment}
\usepackage[colorlinks,citecolor=blue,urlcolor=blue,linkcolor=blue]{hyperref}
\usepackage{ulem}

\newcommand {\VAH}   {{\tt VAH}}
\newcommand {\VH}   {{\tt VH}}

\newcommand{\trento}{T\raisebox{-0.5ex}{R}ENTo}
\newcommand{\trenton}{T\raisebox{-0.5ex}{R}ENTo-3D}

\renewcommand{\P}{\mathcal{P}}
\newcommand{\E}{\mathcal{E}}

\begin{document}

\title{Exploring the fluid behavior in p+p collisions at $\sqrt{s}=13\rm\ TeV$ with viscous anisotropic hydrodynamics}

\author{Shujun Zhao}
\affiliation{School of Physics, Peking University, Beijing 100871, China}

\author{Yiyang Peng}
\affiliation{School of Physics, Peking University, Beijing 100871, China}

\author{Ulrich Heinz}
\affiliation{Department of Physics, The Ohio State University, Columbus, OH 43210-1117, USA}

\author{Huichao Song}
\email{Huichao Song: huichaosong@pku.edu.cn}
\affiliation{School of Physics, Peking University, Beijing 100871, China}
\affiliation{Center for High Energy Physics, Peking University, Beijing 100871, China}
\date{\today}

\begin{abstract}
The applicability of hydrodynamics in small collision systems remains controversial due to the small size and short lifetime of the system. In this letter, we employ viscous anisotropic hydrodynamics (\VAH), which incorporates large pressure anisotropies, to study the collectivity in p+p collisions at $\sqrt{s}=13\ \mathrm{TeV}$. \VAH\ provides a good description for $v_{2}\{2\}$ and $v_{3}\{2\}$ over a wide range of multiplicities and correctly reproduces the experimentally observed negative $c_{2}\{4\}$. Traditional second-order viscous hydrodynamics (\VH), on the other hand, can describe the measurements, in particular the negative $c_{2}\{4\}$, only with model parameters for which the bulk of the evolution is characterized by large values of the shear Knudsen number. It also can not capture the large longitudinal/transverse pressure anisotropy during the early evolution. These demonstrate the failure of traditional viscous hydrodynamics in small collision systems and establish viscous anisotropic hydrodynamics as a more reliable framework to describe the bulk evolution and the observed anisotropic flow in p–p collisions at the LHC.
\end{abstract}

\maketitle

{\em {\it Introduction.}} Relativistic heavy-ion collisions at the Relativistic Heavy Ion Collider (RHIC) and the Large Hadron Collider (LHC) aim to create and study the Quark-Gluon Plasma (QGP), a form of hot QCD matter that once filled the entire universe. Various observables, including transverse momentum, anisotropic flow and their correlators, suggest that the created QGP is strongly coupled and features strong collectivity~\cite{BRAHMS:2004adc, PHOBOS:2004zne, STAR:2005gfr, PHENIX:2004vcz, Gyulassy:2004vg, Muller:2006ee, 796947, Jacobs:2007dw, ALICE:2022wpn}. The quantitative description of these soft observables within the framework of viscous hydrodynamics further suggests that the QGP is a nearly perfect liquid with a very small specific shear viscosity close to the KSS bound predicted by the AdS/CFT correspondence~\cite{Heinz:2013th, Gale:2013da, Song:2017wtw, Song:2013gia,  Bernhard:2019bmu, JETSCAPE:2020shq}.

In recent years collectivity has also been observed in small collision systems, such as p+p, p+Pb collisions at the LHC and $\mathrm{p/d/^3He}$+Au collisions at RHIC. Evidence includes the longitudinal long-range structure in two-particle correlations~\cite{CMS:2012qk, ATLAS:2014qaj, ALICE:2012eyl}, the mass ordering~\cite{ALICE:2013snk, CMS:2014und, Zhao:2017rgg, Weller:2017tsr, Bozek:2013uha},  and valence quark scaling~\cite{Zhao:2020wcd, YuanyuanWang:2023wbz, ALICE:2024vzv} of elliptic flows, and the strangeness enhancement~\cite{ALICE:2016fzo, Kanakubo:2019ogh}, all of which indicate the formation of the QGP droplets. However, traditional viscous hydrodynamics fails to describe even qualitatively some key measurements, most notably the changing sign of the four-particle cumulant $c_{2}\{4\}$ which is negative in high-multiplicity p+p collisions—the so-called “sign puzzle”~\cite{ALICE:2014dwt, CMS:2015yux, ATLAS:2013jmi, Zhao:2017rgg,Zhao:2020pty, Schenke:2019pmk}. This discrepancy raises fundamental questions about the validity of a hydrodynamic description and the origin of collectivity in small systems.

In small collision systems, smaller system sizes and shorter lifetimes result in larger deviations from local equilibrium, which also lead to larger pressure anisotropies, challenging the traditional hydrodynamic approach especially during the early evolution. To address these limitations and extend the applicability of hydrodynamics, various approaches have been developed to incorporate viscous effects non-perturbatively \cite{Bazow:2013ifa, Bazow:2015cha, Tinti:2015xwa, Molnar:2016vvu,Chesler:2015bba, Heller:2015dha,  McNelis:2018jho, Kurkela:2018wud, McNelis:2021zji, Chattopadhyay:2023hpd, Chiu:2025mau}. Among these, viscous anisotropic hydrodynamics (\VAH) offers a promising way to describe systems with anisotropic expansion \cite{Bazow:2013ifa, Bazow:2015cha, Tinti:2015xwa, Molnar:2016vvu, McNelis:2018jho, McNelis:2021zji}. \VAH\ specifically addresses  the large longitudinal/transverse pressure anisotropy, $\P_L/\P_\perp$, generated by the large longitudinal expansion rate during the earliest stage of heavy-ion collisions and treats it non-perturbatively, while maintaining a linearized treatment of the much smaller residual components of the shear stress tensor. The validity of \VAH\ relies on the smallness of the latter while traditional viscous hydrodynamics (\VH) is limited by the much larger $\P_L/\P_\perp$ ratio \cite{McNelis:2021zji, zhao25}.

In this Letter we use \VAH\ to evaluate the fluid behavior of the small fireballs created in p+p collisions at $\sqrt{s}=13\rm\ TeV$. We will show that \VAH\ can be tuned to provide a good description for $v_{2}\{2\}$ and $v_{3}\{2\}$ over a wide range of multiplicities and to reproduce the experimentally measured negative $c_{2}\{4\}$ in high multiplicity events. On the other hand, the description of the measured flow coefficients flow with \VH\ necessarily requires running the code mostly in domains with large Knudsen number where its results are unreliable. Our study demonstrates that \VAH, with its separate evolution of $\P_L$ and $\P_\perp$, provides a more robust description of the collective expansion in p+p collisions than \VH, highlighting its unique utility to study collectivity in small collision systems at the LHC.

{\em {\it Model setup.}}\label{sec:2}
Our simulations are done with a hybrid code that couples a fluid dynamical description of the QGP (with \VAH\ or \VH, for comparison) with the hadronic cascade \cite{SMASH:2016zqf} for the description of the late hadronic rescattering and freeze-out stage. Focusing on measurements at midrapidity, we limit the computational complexity by implementing both \VAH\ and \VH\ in the (2+1)-dimensional event-by-event simulation mode with longitudinal boost invariance. A detailed description of the model setup will be published in \cite{zhao25} --- here we only give a bare-bones summary to put our results in perspective.

In \VAH\ the energy-momentum tensor is decomposed as \cite{Molnar:2016vvu, McNelis:2018jho}
  \begin{align}\label{eq:vahtmn}
    T^{\mu\nu}=\E u^\mu u^\nu+\P_Lz^\mu z^\nu-\P_\perp \Xi^{\mu\nu}
                      +2W_{\perp z}^{(\mu}z^{\nu)}+.\pi_\perp^{\mu\nu},
  \end{align}
where $u^\mu$ is the time-like flow velocity in the Landau frame, $z^\mu$ is a space-like vector defining the direction of the momentum anisotropy, and $\Xi^{\mu\nu}=g^{\mu\nu}-u^\mu u^\nu+z^\mu z^\nu$ is the projection tensor orthogonal to both $u^\mu$ and $z^\mu$. $\E$ is the local energy density, $\P_L$ and $\P_\perp$ are the longitudinal and transverse pressures, and $\pi_\perp^{\mu\nu}$ and  $W_{\perp z}^\mu$ are the transverse shear stress tensor and longitudinal momentum diffusion current.  The term $W_{\perp z}^{(\mu}z^{\nu)}$ denotes the symmetrization of $W_{\perp z}^\mu z^\nu$, which vanishes in $(2+1)$-dimensional case.

The energy momentum tensor $T^{\mu\nu}$ satisfies the conservation law $\partial_\mu T^{\mu\nu}=0$. The additional evolution equations for the anisotropic and diffusive quantities, $\P_L$, $\P_\perp$, $\pi_\perp^{\mu\nu}$ and $W_{\perp z}^\mu$, can be derived from the relativistic Boltzmann equation with an anisotropic particle distribution, $f_a(x,p)=f_{\rm eq}\left(\frac{\sqrt{\Omega_{\mu\nu}p^\mu p^\nu}}{\Lambda(x)}\right)$, using the $14$-moments approximation~\cite{McNelis:2018jho, zhao25}. Here $f_{\rm eq}$ is the equilibrium distribution, $\Lambda(x)$ is the effective local temperature,  $\Omega_{\mu\nu}p^\mu p^\nu=m^2+\frac{p_{\perp,\rm LRF}^2}{\alpha_\perp^2}+\frac{p_{z,\rm LRF}^2}{\alpha_L^2}$, with two parameters $\alpha_{\perp}$ and $\alpha_{L}$ associated with the anisotropy in the transverse and longitudinal directions.

The Equation of State (EoS) is an essential input for the hydrodynamic simulations. \VH\ directly inputs the lattice QCD equation of state with zero chemical potential~\cite{HotQCD:2014kol}. For \VAH, a quasi-particle EoS with effective mass $m(T)$ ~\cite{Molnar:2016vvu, McNelis:2018jho, zhao25} is further constructed to fit this lattice QCD EoS~\cite{HotQCD:2014kol}, together with an evolving mean field $B$ to maintain thermodynamic consistency. The mean field $B$ also modifies the dynamical variables $(\E,\ \P_L,\ \P_\perp)$. Using these modified variables, the anisotropic parameters $(\Lambda,\alpha_L,\alpha_\perp)$ are determined through the generalized Landau matching condition for each fluid cell and each evolution time step. Then the anisotropic particle distribution $f_a$ is obtained, and the anisotropic transport coefficients in the evolution equations for $\P_L,\ \P_\perp,\ \pi^{\mu\nu}$, and $W_{\perp,z}^\mu$ can be calculated from the kinetic theory. The shear and bulk relaxation times $\tau_{\pi}$ and $\tau_{\Pi}$ in \VAH\ are phenomenological inputs and assumed to take the same form as used in \VH~\cite{Denicol:2012cn}: $\tau_{\pi}=\eta/\beta_\pi$, $\tau_\Pi=\zeta/\beta_\Pi$, with $\beta_{\pi}$ and $\beta_{\Pi}$ calculated in Refs.~\cite{Jaiswal:2021uvv, Molnar:2016vvu, McNelis:2018jho}. For the specific shear and bulk viscosities $\eta/s$ and $\zeta/s$, we adopt the same temperature dependent parametrizations as proposed by the JETSCAPE Collaboration~\cite{JETSCAPE:2020mzn}, in both \VH\ and \VAH.

\begin{table}[t]
    \centering
    \caption{Main parameters in \trenton\ that influence multiplicity fluctuations and flow predictions for \VAH\ and \VH\ simulations.}
    \begin{tabular}{|c|c|c|c|c|c|c|c|c} 
        \hline
        model & para & $k_\beta$& $w$ (fm) & $v$ (fm) & $n_c$\\
        \hline
        \multirow{3}{*}{\VAH} & I   & $0.30$ & $0.90$ & $0.52$ & $6$ \\
                              & II  & $0.33$ & $0.90$ & $0.52$ & $6$ \\
                              & III & $0.70$ & $0.90$ & $0.50$ & $5$ \\
        \hline
        \multirow{3}{*}{\VH}  & IV  & $0.19$ & $0.80$ & $0.30$ & $7$ \\
                              & V   & $0.19$ & $0.80$ & $0.30$ & $7$ \\
                              & VI  & $0.28$ & $0.90$ & $0.47$ & $3$ \\
        \hline
    \end{tabular}
    \label{tab:inipara}
\end{table}

\begin{table}[t]
       \centering
    \caption{Parameters for $(\eta/s)(T)$ for \VAH\ and \VH\ simulations.}
    \begin{tabular}{|c|c|c|c|c|c|} 
        \hline
        model & para & $(\eta/s)_{\min}$ & $T_\eta$ (GeV) & $\ \ a_{\rm low}\ \ $ & $a_{\rm high}$\\
        \hline
        \multirow{3}{*}{\VAH} & I   & $0.08$ & $0.146$ & $\ -0.383\ $ & $\ 0.393\ $\\
                              & II  & $0.11$ & $0.146$ & $-0.383$ & $0.393$\\
                              & III & $0.14$ & $0.146$ & $-0.383$ & $0.393$\\
        \hline
        \multirow{3}{*}{\VH}  & IV  & $0.08$ & $0.146$ & $-0.383$ & $0.393$\\
                              & V   & $0.12$ & $0.146$ & $-0.383$ & $0.393$\\
                              & VI  & $0.28$ & $0.146$ & $0$ & $0$\\
        \hline
    \end{tabular}
    \label{tab:hydropara}
\end{table}

\begin{figure*}[tbh]
    \centering
        \includegraphics[width=0.42\textwidth]{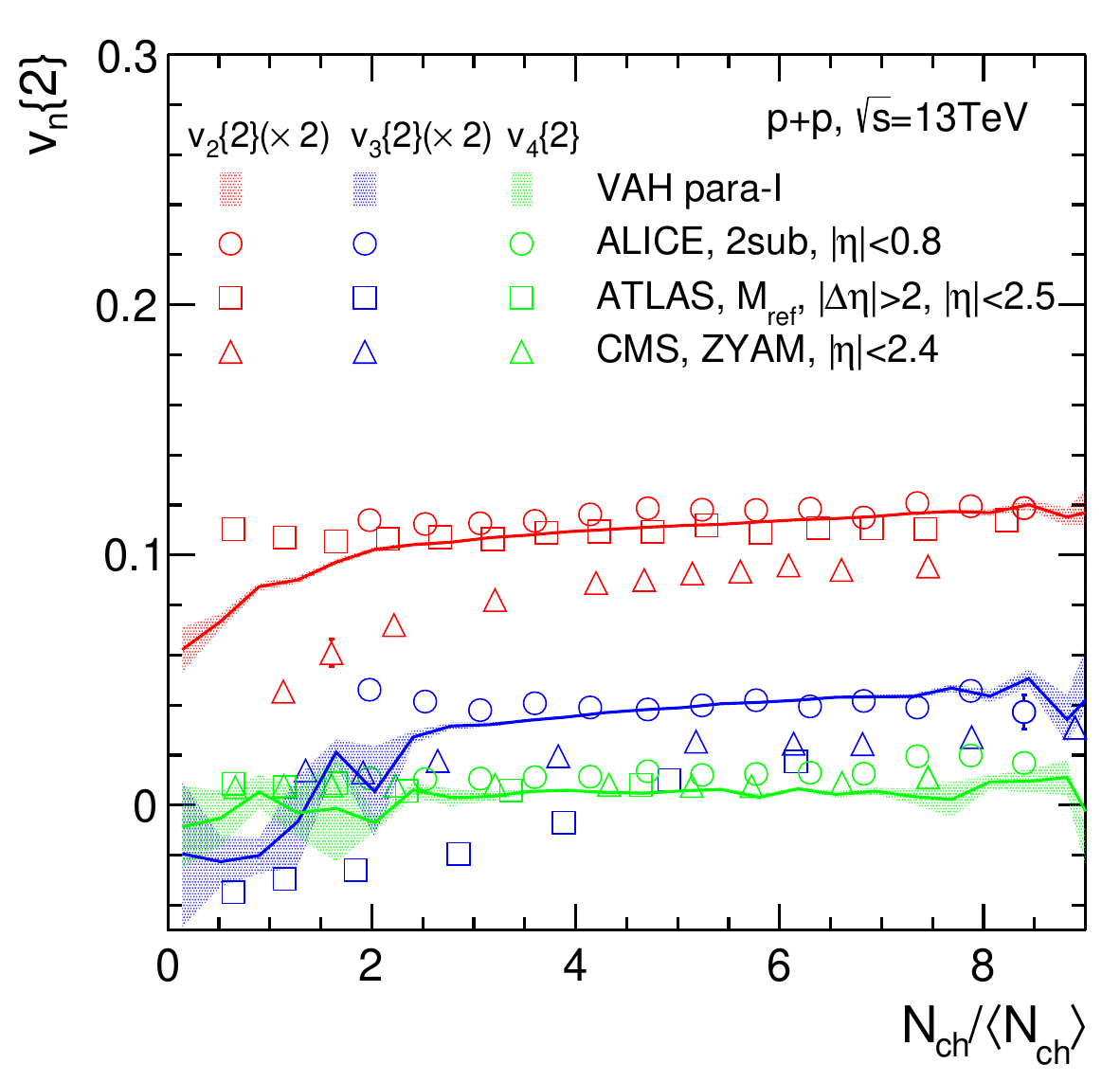}
    \hspace{0.5cm}
        \includegraphics[width=0.42\textwidth]{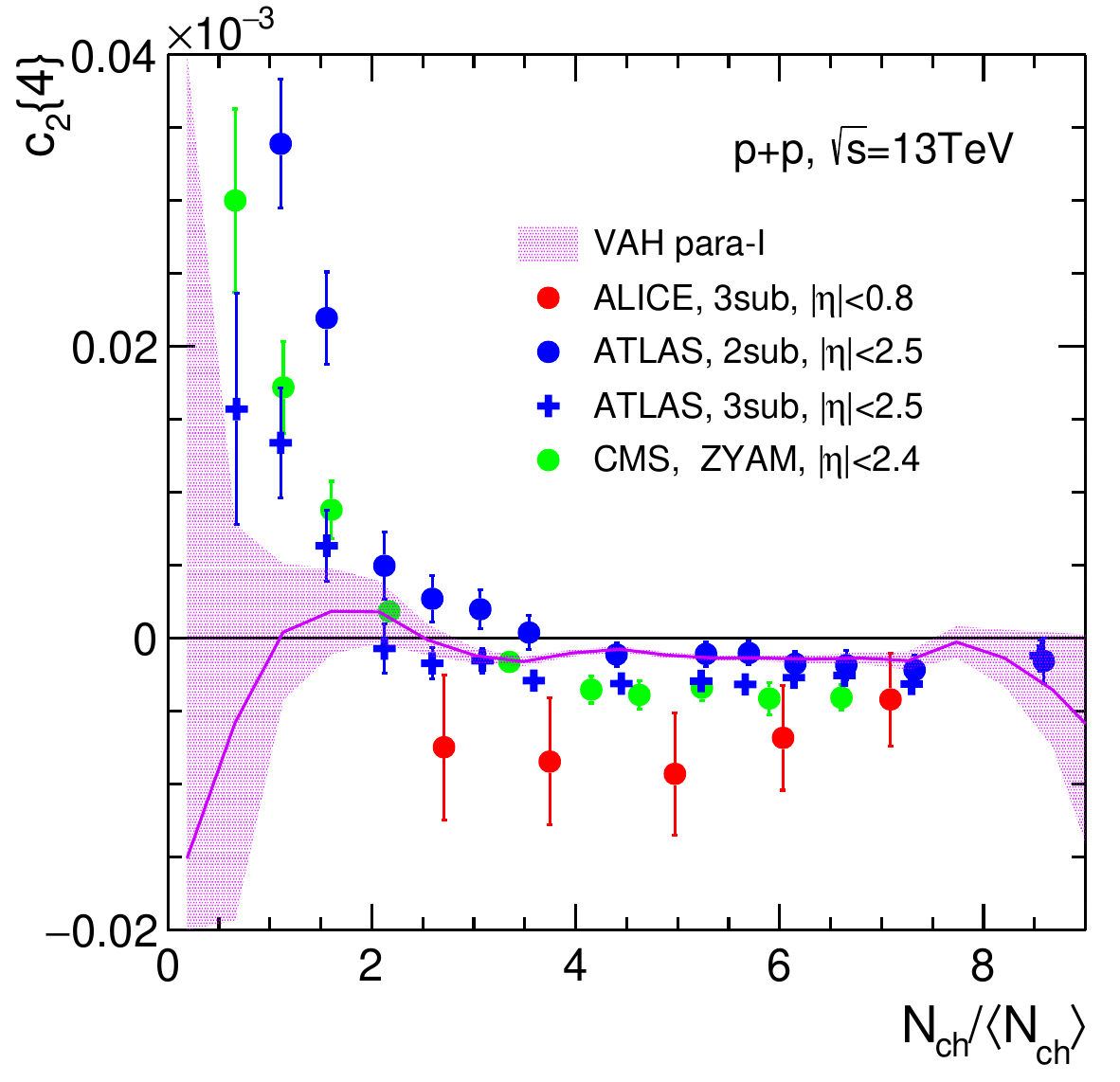}
    \caption{Multiplicity dependent $v_{n}\{2\}$ ($n=2,3,4$) (left) and $c_2\{4\}$ 
    (right) in p+p collisions at $\sqrt{s}=13\rm\ TeV$, calculated from event-by-event viscous anisotropic hydrodynamic model (\VAH) with parameter-sets para-I, together with a comparison to the ALICE~\cite{ALICE:2019zfl}, ATLAS~\cite{ATLAS:2017hap, ATLAS:2017rtr} and CMS~\cite{CMS:2016fnw, CMS:2017kcs} data.\label{fig:flowvah}}
\end{figure*}

Near the quark-hadron phase transition the hydrodynamic fluid cells convert into hadrons on a switching (``particlization'') hypersurface of constant temperature $T_{\rm sw}$, using a particle event generator based on the Cooper-Frye formula~\cite{Cooper:1974mv} with specifically chosen distribution functions. For \VAH\ we adopt the anisotropic PTMA distribution \cite{McNelis:2019auj, McNelis:2021acu} while for \VH\ we use the isotropic PTB distribution ~\cite{Pratt:2010jt, McNelis:2021acu}. The emitted hadrons are then fed into the SMASH hadron cascade model \cite{SMASH:2016zqf} for subsequent elastic and inelastic scatterings and resonance decays until final freeze-out.

The hydrodynamic simulations start at a fixed proper time ($\tau_0=0.12\ \rm fm/c$ for \VAH, and $\tau_0=0.6\ \rm fm/c$ for \VH) with zero initial flow velocity. The initial profiles are generated from the parameterized 3-dimensional \trenton\ model with sub-nucleon fluctuations \cite{Soeder:2023vdn}. For the $(2{+}1)$-D simulations with longitudinal boost invariance performed in this paper, the initial energy densities are obtained from the \trenton\ profiles at space-time rapidity $\eta_s=0$. The initial pressure anisotropy for \VAH\ is set to $(\P_L/\P_\perp)_0=0.4$. For both \VAH\ and \VH\ the particlization temperature is set to $T_{\rm sw}=146\rm\ MeV$. The $14$ free parameters in \trenton, together with the parameters for $(\eta/s)(T)$ and $(\zeta/s)(T)$, are tuned to roughly fit the particle yields, multiplicity distributions $P(N_{\rm ch})$, and the flow harmonics $v_2\{2\}$ and $v_3\{2\}$ for all charged hadrons in p+p collisions at $\sqrt{s}=13\rm\ TeV$ measured by ALICE~\cite{ALICE:2022xip, ALICE:2019zfl} (para I and para IV in Tables~\ref{tab:inipara} and \ref{tab:hydropara}). In fact the specific shear viscosities $(\eta/s)(T)$ in para I and para IV are very close to the one inferred by the Bayesian calibration of \VAH\ with Pb+Pb collisions at the LHC~\cite{Liyanage:2023nds}. For the specific bulk viscosity $(\zeta/s)(T)$ we shift its peak from 230 MeV~\cite{Liyanage:2023nds} to 150 MeV, which we checked has no large influence on the  predicted flow observables, but brings it closer to the Bayesian calibration of \VH\ by the JETSCAPE collaboration \cite{JETSCAPE:2020mzn}. To systematically study fluid behavior in p+p collisions we also explore the model's predictions for several other parameter sets (para II, III for \VAH\ and para V, VI for \VH\ in Tables~\ref{tab:inipara} and \ref{tab:hydropara} as well as for different regulation schemes.\footnote{%
	Both \VH\ and \VAH\ implement specific regulation schemes to suppress numerical instabilities. 
	For \VH, reg1 reduces the magnitudes of $\pi^{\mu\nu}$ by a suppression factor 
	which ensures the transversality and tracelessness conditions, while reg2 only 
	reduces the magnitudes of $\pi^{\mu\nu}$. For \VAH\ only the transverse components of the 
	shear stress, $\pi_\perp^{\mu\nu}$, are regulated, by a factor that preserves 
	its transversality and tracelessness. {To ensure reliable numerical simulations, these regulations are required to be properly implemented without affecting the QGP evolution within $T_c$~\cite{zhao25}. }
}

Tables~\ref{tab:inipara} and \ref{tab:hydropara} list the key parameters. Among these, the initial state parameter $k_\beta$ primarily influences the multiplicity distribution, while the nucleon width $w$, constituent width $v$, and constituent number $n_c$ affect both the flow and multiplicity distributions. $(\eta/s)(T)$ is modeled by a piecewise linear form with slopes $a_L$ and $a_H$ intersecting at $T_\eta$, bounded from below by $(\eta/s)_{\rm min}$. For details on the models, different setups, as well as the regulation procedures for \VAH\ and \VH\, please refer to Refs.~\cite{zhao25, McNelis:2021zji}.

{\em {\it Results and discussion.}}\label{sec:3}
Figure~\ref{fig:flowvah} shows the flow harmonics  $v_{n}\{2\}$ ($n=2,3,4$) and the 4-particle cumulant  $c_2\{4\}$ as a function of the normalized multiplicity $N_{\rm ch}/ \langle N_{\rm ch}\rangle $  in p+p collisions at $\sqrt{s}=13\rm\ TeV$, calculated from \VAH\ (para I) and measured in experiments. The differences among the measured $v_{n}\{2\}$ ($n=2,3,4$) arise mainly from different kinematic cuts and non-flow subtraction methods used by the three experiments (ALICE, ATLAS and CMS). For the model predictions we implement the kinematic cut $|\eta| < 2.0$ and $0.2 < p_T < 3\ \rm{GeV}/c$) and subtract non-flow effects using the two-subevent method with a finite $\eta$-gap $\Delta \eta>0.2$. The model parameters (para I) were tuned to fit the ALICE measurements \cite{ALICE:2022xip, ALICE:2019zfl} of the multiplicity distribution $P(N_{\rm ch})$ and the flow anisotropies $v_2\{2\}$ and $v_3\{2\}$ in the range $4{\,<\,}N_{\rm ch}/\langle N_{\rm ch}\rangle{\,<\,}8$. Fig.~\ref{fig:flowvah} shows that, with this tuning, \VAH\ slightly underpredicts the ALICE, ATLAS and CMS measurements of the 4th flow harmonic $v_4\{2\}$ while correctly reproducing the negative sign of the 4-particle cumulant $c_2\{4\}$ at the high end of the multiplicity range ({\it i.e.} $3{\,<\,}N_{\rm ch}/\langle N_{\rm ch}\rangle{\,<\,}8$ which is consistently reported by all three experiments. At these multiplicities, the predicted magnitude of $c_2\{4\}$ roughly agrees with the ATLAS data obtained with the two-subevent non-flow subtraction method whereas the CMS and ALICE collaborations report negative values of somewhat larger magnitude (albeit with much larger uncertainties in the latter case). While this is not the place for an in-depth analysis of the differences between these experimental measurements, we highlight that our \VAH\ simulation is the first hydrodynamic model that correctly reproduces the negative sign of $c_2\{4\}$ that is consistently reported by the experiments in high-multiplicity p+p events. Multiple previous attempts using traditional viscous hydrodynamics, whether with HIJING, AMPT, \trento, or IP-Glasma initial conditions \cite{Zhao:2017rgg, Zhao:2020pty, Schenke:2019pmk}, all failed to do so. As will be discussed in Ref.~ \cite{zhao252}, the same \VAH\ simulations shown in Fig.~\ref{fig:flowvah} also provide qualitatively accurate descriptions for other flow observables in high multiplicity p+p collisions, including the 6-particle cumulant $c_2\{6\}$, the symmetric cumulants $\rm{SC}(2,3)$ and $\rm{SC}(2,4)$, and the asymmetric cumulants $\rm{ac}_{2}\{3\}$, and others.

\begin{figure*}[tbh]
    \centering
        \includegraphics[width=0.42\textwidth]{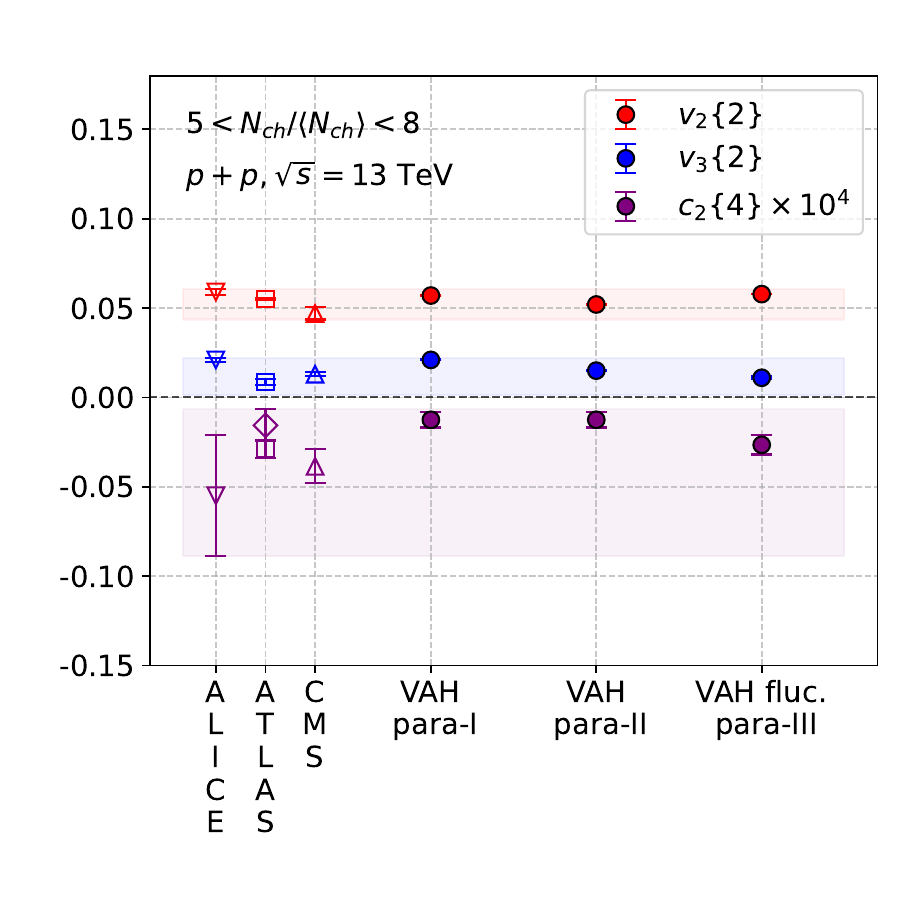}
    \hspace{0.5cm}
        \includegraphics[width=0.42\textwidth]{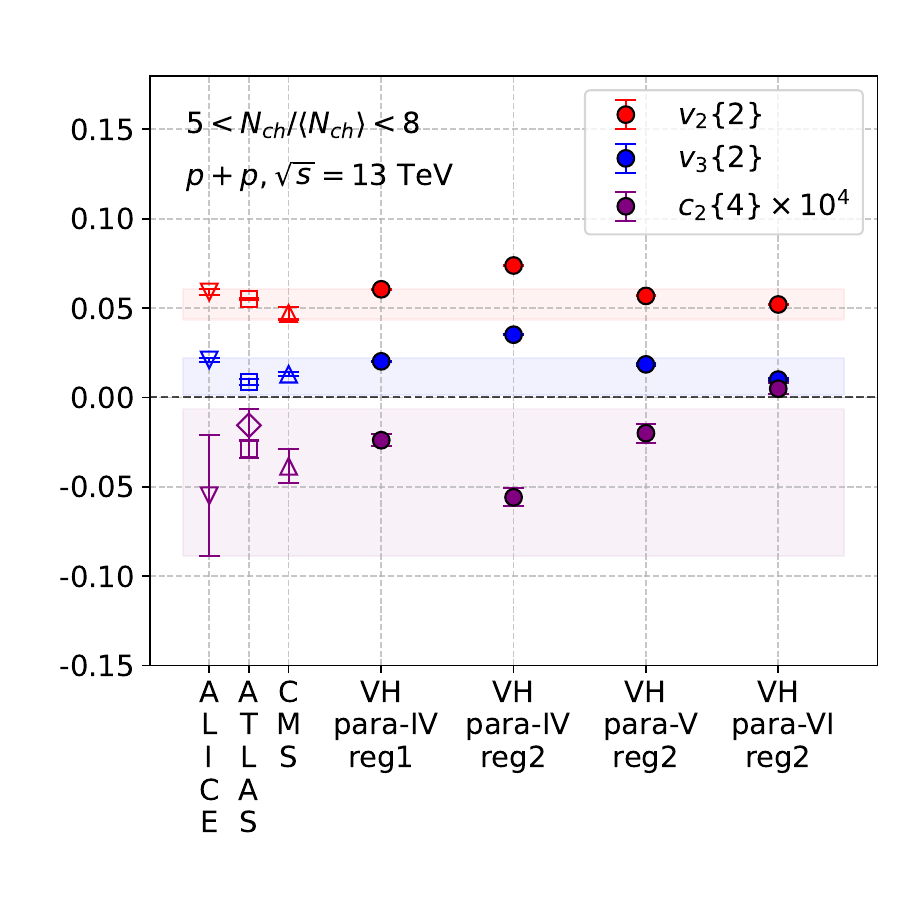}
        \caption{The integrated $v_{2}\{2\}$, $v_{3}\{2\}$ and $c_2\{4\}$ within 
        $5{\,<\,}N_{\rm ch}/\langle N_{\rm ch}\rangle{\,<\,}8$ in p+p collisions at 
        $\sqrt{s}=13\rm\ TeV$, calculated from viscous anisotropic hydrodynamics
        (\VAH, left panel)  and traditional viscous hydrodynamics (\VH, right panel) 
        with different parameter sets and regulation schemes (see text). 
        Experimental data are taken from the ALICE~\cite{ALICE:2019zfl}, 
        ATLAS (diamonds and squares denote two- and three-subevent $c_2\{4\}$, 
        respectively) \cite{ATLAS:2017hap, ATLAS:2017rtr} and CMS
        \cite{CMS:2016fnw, CMS:2017kcs} collaborations. The red, blue, and purple 
        shaded horizontal bands correspond to the upper and lower limits of the
        $v_{2}\{2\}$, $v_{3}\{2\}$ and $c_2\{4\}$ data measured by these three 
        collaborations, respectively.
    \label{fig:vah}}
\end{figure*}

\begin{figure*}[tbh]
    \centering
        \includegraphics[width=0.4\textwidth]{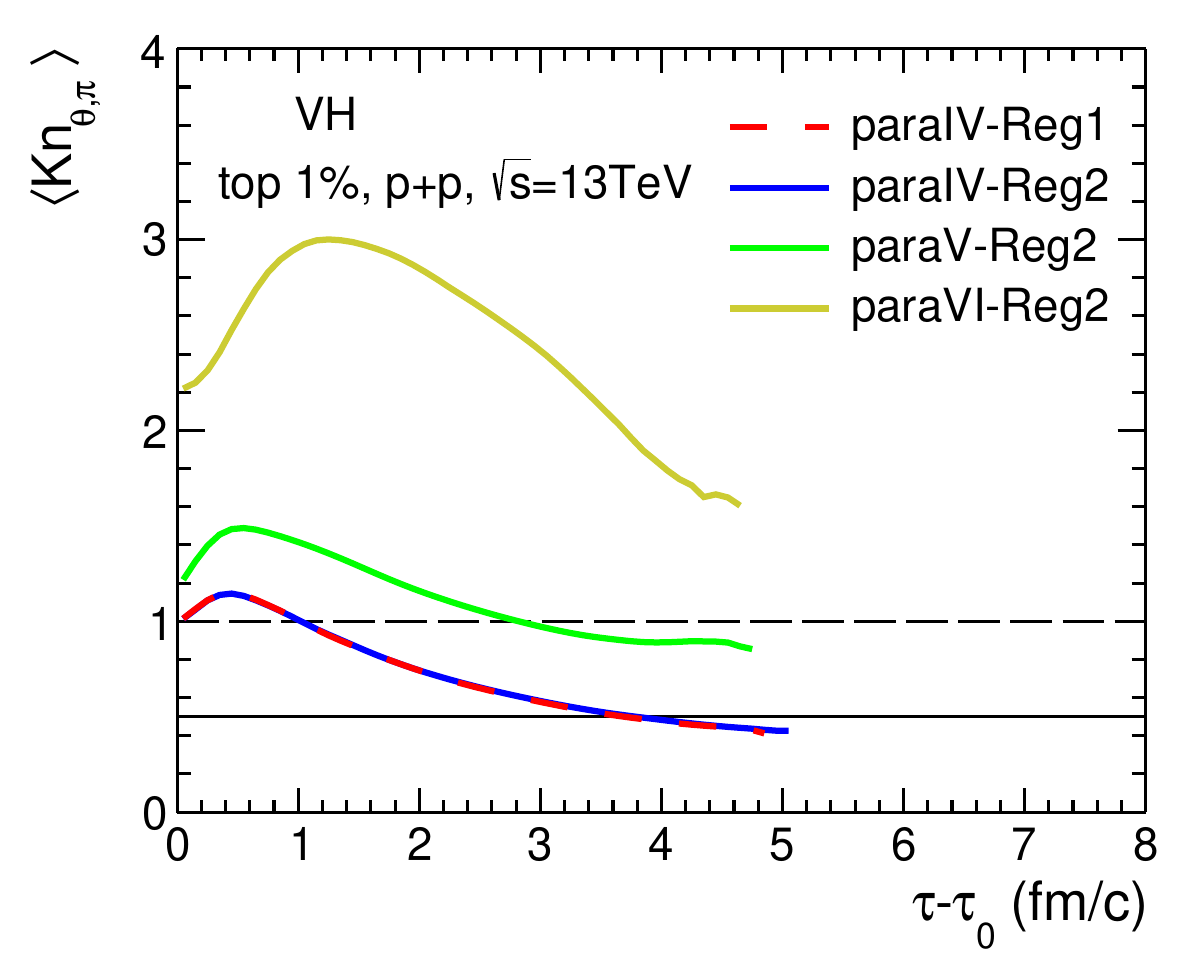}
    \hspace{0.5cm}
       \includegraphics[width=0.42\textwidth]{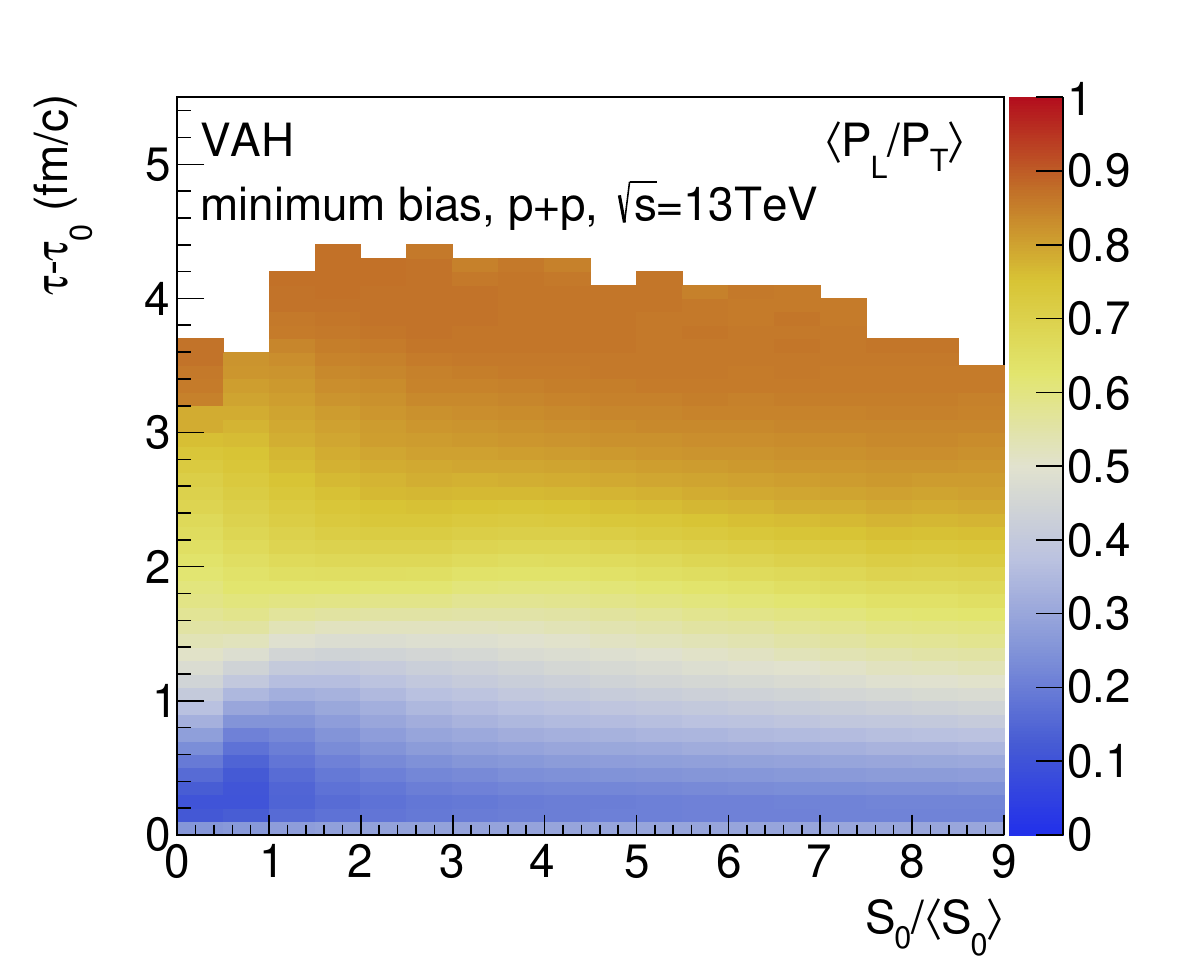}
    \caption{Time evolution of the average Knudsen number $\langle \text{Kn}_{\theta,\pi}\rangle$ 
    from \VH\ simulations (left) and the average pressure anisotropy $\langle\P_L/\P_T\rangle$ 
    from \VAH\ simulations (right) in p+p collisions at $\sqrt{s}=13$ TeV. The average is taken 
    within the fireball ($\varepsilon>\varepsilon_{\rm sw}$) and weighted by the local energy density. \label{fig:reg}}
\end{figure*}

In Fig.~\ref{fig:vah} we explore the sensitivity of the integrated $v_{2}\{2\}$, $v_{3}\{2\}$ and $c_2\{4\}$ to the choice of hydrodynamic model (\VAH\ vs. \VH), for different sets of model parameters and regulation schemes. The left panel shows that for \VAH, replacing parameter set para I by para II which increases the specific shear viscosity $\eta/s$, leads to a slight decrease of $v_{2}\{2\}$ and $v_{3}\{2\}$ while $c_2\{4\}$ remains negative. Parameter set para III increases the sub-nucleonic  fluctuations in the initially deposited energy profiles deposition (see Ref.~\cite{zhao25} for details) but still yields a negative $c_2\{4\}$. The right panel of Fig.~\ref{fig:vah} demonstrates that, with fine-tuned model parameters, traditional viscous hydrodynamics \VH\ can also reproduce a negative $c_2\{4\}$ and roughly describe the measured $v_{2}\{2\}$ and $v_{3}\{2\}$. In this case, however, this success comes at an unacceptably high cost: as already mentioned, both \VAH\ and \VH\ need some regulation procedures to stabilize the numerical simulations, especially for event-by-event simulations with fluctuating initial conditions. For the \VAH\ simulations reported here, only {a portion} of rather cold fluid cells far outside the QGP require {large} regulation of $\pi_\perp^{\mu\nu}$, and {the influence on the hot QGP fireball evolution within $T_c$ is negligible}~\cite{zhao25}, so it will not influence the hadrons emitted from the particlization surface. On the other hand, in our phenomenologically successful \VH\ runs we find a much larger number of fluid cells in need of regulation of $\pi^{\mu\nu}$, and (especially for the reg2 prescription) strong regulation happens in a sufficiently large space-time volume {\it within} the QGP medium \cite{zhao25} that final hadronic observables are significantly affected (see the columns for \VH\ para IV with regulations reg1 and reg2 in the right panel). It is thus no longer clear that the regulated code \VH\ solves the correct fluid dynamic equations of motion with the desired values for the specific viscosities.   

The limitations of traditional viscous hydrodynamics \VH\ are further illustrated in the left panel of  Fig.~\ref{fig:reg} which plots the time evolution of the average Knudsen number $\langle \text{Kn}_{\theta,\pi} \rangle$ within the fireball obtained from the \VH\ simulations. Here $\text{Kn}_{\theta,\pi} \equiv \tau_\pi \theta$ and $\theta=\partial_\mu u^\mu$ is the local expansion rate~\cite{Niemi:2014wta}.  The validity of traditional viscous hydrodynamics requires that the relaxation time $\tau_\pi$ is much smaller than the inverse of the local expansion rate, $\tau_\pi{\,\ll\,}1/\theta$, in order to maintain approximate local equilibrium during the evolution. However, the parameter sets used in the phenomenologically successful \VH\ simulations that roughly fit $v_{2}\{2\}$ and $v_{3}\{2\}$ and reproduce a negative $c_2\{4\}$ in p+p collisions all exhibit quite large average Knudsen numbers, $\langle \text{Kn}_{\theta,\pi} \rangle{\,>\,}1$, especially during the early time evolution.\footnote{%
	Note that in \VH\ with parameters para IV the evolution of the dissipative terms
	is controlled by the {\it same} Knudsen number irrespective of the regulation scheme 
	selected (solid blue and dashed red curves in the left panel of Fig.~\ref{fig:reg}) but 
	that the resulting shear stress tensor $\pi^{\mu\nu}$ and final flow observables 
	(especially the fragile 4-particle cumulant $c_2\{4\}$) are quite different for the 
	two regulation schemes reg1 and reg2 (right panel of Fig.~\ref{fig:vah}). This 
	severe regulation scheme dependence also renders \VH\ predictions for p+p collisions 
	at the LHC unreliable.}
This directly demonstrates the failure of traditional viscous hydrodynamics for small collision systems.  

Finally the right panel of Fig.~\ref{fig:reg} shows for parameter set para I the \VAH\ time evolution of the average pressure anisotropy $\langle\P_L/\P_T\rangle$ inside the QGP fireball in p+p collisions at the LHC. The horizontal axis uses the normalized total initial entropy $S_0/ \langle S_0 \rangle$, where $\langle S_0\rangle$ is the event-averaged value, as a proxy for the normalized final charged multiplicity density $N_{\rm ch}/\langle N_{\rm ch}\rangle$. For all values of $S_0/ \langle S_0 \rangle$ the system exhibits $\langle\P_L/\P_T\rangle{\,<\,}1$, {\it i.e.} the QGP fluid does not isotropize before it hadronizes. Remembering that traditional viscous hydrodynamics \VH\ is based on the assumption of small pressure anisotropies throughout the evolution, this panel provides another direct demonstration of the failure of \VH\ in small collision systems.

{\em {\it Summary and outlook. }}\label{sec:4}
In this paper we implement viscous anisotropic hydrodynamics (\VAH) with \trenton\ initial conditions to study flow and evaluate the fluid behavior of p+p collisions at $\sqrt{s}=13$\,TeV. \VAH\ extends traditional viscous hydrodynamics by treating large dissipative corrections arising from large differences between the longitudinal and transverse expansion rates non-perturbatively. With properly tuned parameters, \VAH\ provides a decent description of the $p_T$-integrated $v_{2}\{2\}$ and $v_{3}\{2\}$ data over a wide range of multiplicities. It also correctly predicts a negative $c_{2}\{4\}$ that qualitatively agrees with the experimental measurements at high multiplicities. Although traditional viscous hydrodynamics (\VH) can also roughly describe the flow data and generate a negative $c_{2}\{4\}$ with tuned model parameters, these simulations operate the model mostly in domains featuring large Knudsen numbers, require serious numerical regulation which affects the predicted values for the flow observables, and are thus unreliable and cannot be trusted for quantitative analysis. \VAH\ evolves the longitudinal and transverse pressures $\P_L$ and $\P_\perp$ independently, thereby accounting for strong pressure anisotropies non-perturbatively, and yields predictions that are not affected by the numerical regulation prescription. This makes \VAH\ a highly preferred approach for the quantitative analysis of flow and collectivity in p+p collisions and other small collision systems.

{\em {\it Acknowledgements.}} We thank Jiangyong Jia, Weiyao Ke, J{\"u}rgen Schukraft, Wenbin Zhao and You Zhou for useful discussions. This work is supported by the National Natural Science Foundation of China under Grants  No.12575138 and No.12247107. This work was performed at the Dawning Supercomputer Center (Computing Center in Xi'an), the National Supercomputer Center in Tianjin (Tianhe New Generation Supercomputer), and the High-Performance Computing Platform of Peking University.

\bibliography{ref2}

\end{document}